# Ultra-fast synthesis and thermodynamic analysis of MoAlB by self-propagating high temperature combustion synthesis

Hang Yin, Xiaodong He, Guangping Song, Yongdong Yu, Yongting Zheng, Yuelei Bai[*]

National Key Laboratory of Science and Technology on Advanced Composites in Special Environments and Centre for Composite Materials and Structures, Harbin Institute of Technology, Harbin 150080, P. R. China

**Abstract**. MoAlB as a typical member of the MAB phases has attracted much growing attention due to its unique properties. However, the low production of MoAlB powders limits its further development and potential applications. To respond this challenge, the ultra-fast preparation of high-purity MoAlB powders is achieved in a few seconds by self-propagating high temperature combustion synthesis (SHS) using the raw powder mixture at the atomic ratio of 1Mo/1.3Al/1B. The reaction mechanism in SHS process was obtained by analyzing the corresponding composition changes of starting materials. Furthermore, the thermodynamic prediction for the SHS reaction of Mo + (1+ $y$) Al + B = MoAlB + $y$Al is consistent with the present experiments, where the preparation of MoAlB also conforms to two common self-propagating conditions of SHS. Enthalpy vs temperature curve shows that the adiabatic temperature of the reaction decreases with increasing the amount of excuse Al, but increases when pre-heating the reactants. The thermodynamic calculation method also provides a new idea for the preparation of other MAB phases by SHS.

[*] **Corresponding author:** Tel: +86-45186403956; Fax: +86-45186403956; Email: baiyl@hit.edu.cn, baiyl.hit@gmail.com.

1. **Introduction**

Binary borides are highly valued due to their high hardness, high temperature creep resistance, high wear resistance, high melting point, corrosion resistance, and other properties [1]. However, their further application is hindered by some disadvantages, one of which is the intrinsic brittleness. The high fracture toughness and damage tolerance of the MAX phases [2-4] provide an inspiration for this problem by inserting one or two A-group atomic layer(s) into binary borides to form ternary transition metal boride named MAB phase [5] with weak bonding [6], where M is transition metal element, A is IIIA and IVA group elements, and B is boron element [7-9].

Although most of the MAB phases were identified as early as the 1960s [10, 11], the detailed characterization of structure and properties was performed since 2013 when Tan et al [12] revealed the magnetocaloric effect (MCE) of $Fe_2AlB_2$. While recent experimental and theoretical [6, 13] work both confirm that the MAB phases (MoAlB [14] and $Fe_2AlB_2$ [15, 16]) do have high fracture toughness and damage tolerance similar to the MAX phase, originating from their layered structure and weak Al-Al or M-A bonds. Moreover, they have excellent properties such as high strength [16, 17], high elastic moduli [18, 19], electrical conductivity [18, 19], and thermal conductivity [18, 19]. Among the dozens of MAB phases investigated now, MoAlB is undoubtedly the most concerned because as the first fabricated bulk MAB phase, it has been shown high oxidation resistances [20], high-temperature stability [19], crack healing behavior [21] and ablation resistance up to 2050 ºC [22] besides the general features of the MAB phases, which is extremely important for its potential high-temperature structural applications. In addition, the recent interest in MoAlB is also from its 2D derivate, Mbenes [23-25] or Boridene [26, 27].

The breakthrough for processing of MoAlB conducted by Barsoum et al [14] opens the door to understand its intrinsic properties, and explodes the growing research on the MAB phases. Although MoAlB has been synthesized by the conventional processes including the hot pressing [14], pressless sintering [28], spark plasma sintering [29, 30], and thermal explosion [31] at 1050-1200 ºC for 0.25-5.8 h since 2016, these processing methods all suffer from some disadvantages including high cost, high temperature, long time, high energy consumption and complicated process. Furthermore, the long-time heat preservation at high temperature usually results in the impurity oxides such as $Al_2O_3$ as a result of the reaction between Al and residual $O_2$ in the atmosphere.

As one of the frequently-used synthesis processes, the self-propagating high temperature synthesis (SHS) [32] has been widely used for refractory compounds, with the significant advantages of the low cost, simple production process, very short time, little demand on external energy, and convenient commercialization. In this process, the resistance wire and ignition agent are used to ignite a self-propagating reaction among the raw materials because of the highly exothermic reaction, as a result of the synthesis of target materials. Moreover, it has been successfully employed to synthesize some MAX phases, such as $Ti_2AlC$ [33, 34], $Ti_3AlC_2$ [35, 36], $Ti_3SiC_2$ [37], $Ti_2SnC$ [38], and $Nb_2AlC$ [39]. Very recently, Jessica et al. [40] successfully prepared $Mn_2AlB_2$ powders by SHS, indicating its feasibility in MAB phase materials. However, there has been no report on the synthesis of MoAlB by SHS to date.

In the present work, the SHS was employed to synthesize the high-purity MoAlB from commercial powder mixture of Mo, Al, and B as starting materials, where the effect of raw materials on phase composition of products and morphologies are also investigated. In addition, the adiabatic combustion temperature ($T_{ad}$) of the SHS for MoAlB reaction are also presented using available thermodynamic data, providing a prediction basis for the SHS preparation of other MAB phases.

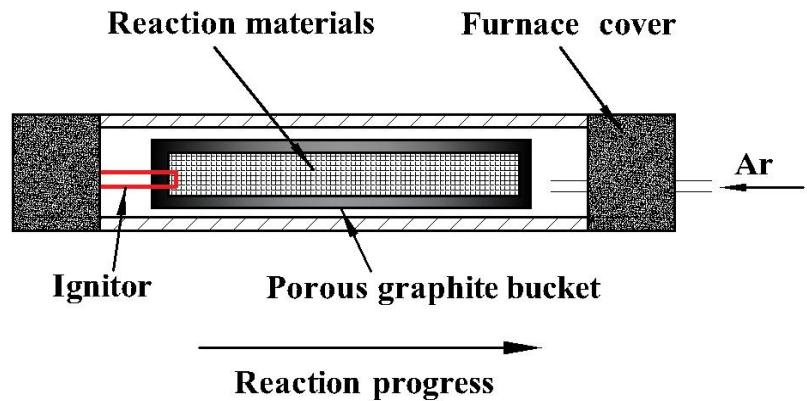

Figure 1 The set-up for self-propagating high-temperature synthesis apparatus.

## 2. Experimental details

*2.1 Synthesis of MoAlB powders by SHS*

In this work, three kinds of Mo powders (10 μm, 99.99%; 1–3 μm, 99.99%; 500nm, 99.95%), three kinds of B powders (5 μm, 99.95%; 1–3 μm, 99.95%; 200 nm, 99.95%), and Al (10 μm, 99.5%) powders were used to prepare MoAlB. Samples containing 20 g of each Mo-Al-B powder mixture are weighed and grounded with one horizontal mixer for 24 h. Then the resulting pellets of this powder mixture were placed on a graphite crucible in apparatus of self-propagating high-temperature synthesis (Figure 1). In Ar, a heated coil at one end ignited the pellet by the spiral Ni-Cr resistance wire to insert the ignition agent (the mixture of Ti and C powders with a molar ratio of 1:1), and the self-sustained combustion wave propagated from this heated end to a full sample because of the highly exothermic reaction. The reaction can be observed as the pressure gauge pointer increased first and then decreased, this process was only a few seconds. Upon cooling to the room temperature, the porous resultants were crushed and grounded with pestle and mortar for further characterization.

*2.2 Thermodynamic calculation details*

In order to exploit the SHS process, the adiabatic combustion temperature ($T_{ad}$) with its dependency on pre-heating temperature and Al diluents is adopted. It is a measure of the maximum temperature reached when the sample is prepared using the combustion synthesis process, assuming no energy loss to the environment, although in practice some heat is always lost.

The thermodynamic study presented here is based on the reaction (1), using available data on MoAlB. Based on the previous study [41], Al as an additive can not only reduce the reaction temperature as a diluent, but also partially improve the purity of the product:

$$\text{Mo} + (1+y)\,\text{Al} + \text{B} = \text{MoAlB} + y\,\text{Al} \tag{1}$$

where $y$ is the excess Al to act as a diluent to compensate for volatilized loss in the process.

The heat release of the combustion reaction determines whether the self-propagation can continue. According to Merzhanov [42], when the adiabatic temperature of the system is greater than 1800 K, the self-propagation can be maintained spontaneously without additional heating, which also corresponds to the condition $-\Delta H_r^{298\,\text{K}}/C_p^{298\,\text{K}} \geqslant 2000\,\text{K}$ [43]. $T_{ad}$ is obtained by taking into account that all the heat released by the reaction is used for the temperature increase of the reaction under the assumption that no heat is lost. For the equilibrium equation in reaction (1) it can be expressed as:

$$-\Delta H_r^{298} = \int_{298}^{T_{ad}} C_p[\text{MoAlB} + y\,\text{Al}]\,dT \tag{2}$$

In addition, the influence of various phase transformation in the reactants and products should also be considered. For example, when the temperature reaches 933 K or higher, Al will melt, which means that the melting enthalpy of Al should be considered:

$$-\Delta H_r^{298} = \int_{298}^{T_{ad}} C_p[\text{MoAlB} + y\,\text{Al}]dT + y\Delta H_m(\text{Al}) \qquad (3)$$

If the reactants are preheated to $T_{start}$ before SHS, the additional energy provided by the initial heating of the reactants must be considered:

$$-\Delta H_r^{298} + \int_{298}^{T_{start}} C_p[\text{Mo} + (1+y)\,\text{Al} + \text{B}]dT = \int_{298}^{T_{ad}} C_p[\text{MoAlB} + y\,\text{Al}]dT \qquad (4)$$

## 3. Results and Discussion

### 3.1 Synthesis and characterization by SHS

#### 3.1.1 Effect of Al molar ratio on formation of MoAlB

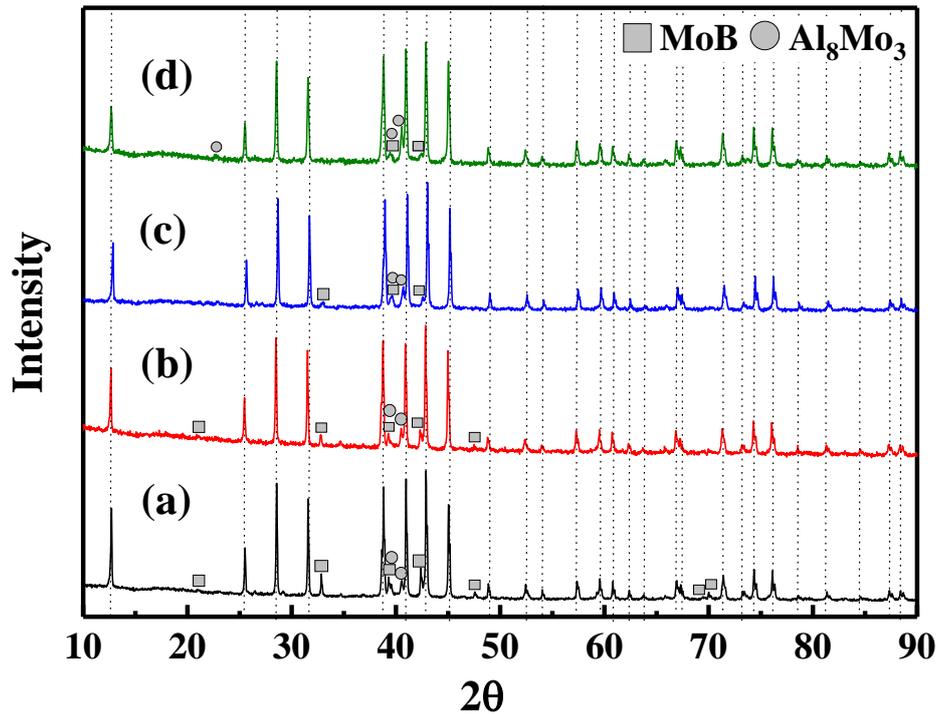

Figure 2 XRD patterns of samples synthesized from 1Mo/$n$Al/1B by SHS: (a) $n$ = 1.1, (b) $n$ = 1.2, (c) $n$ = 1.3, and (d) $n$ = 1.4, where the dotted line represents MoAlB (PDF#65-2497).

In general, excess Al is often added to raw materials in the preparation of MAB phases to compensate for the volatilization of liquid Al [8], so it is very important to determine the proportion of Al in raw powder mixture for the high purity of MoAlB. Samples with Al content from $n$ = 1.1-1.5 were tested. However, because the raw powder mixture cannot be ignited for $n$ = 1.5, the phase composition of the SHSed products is examined by XRD (Figure 2, and Table 1) only for the $n$ = 1.1-1.4. In Figure 2, MoAlB always is the main phase in all cases (MoAlB, PDF#65-2497), with the impurities MoB and $Al_8Mo_3$. With increasing $n$ from 1.1 to 1.3, the content of MoAlB increases, with a significant decreasing amount of MoB impurity (Table 1). Considering the negligible compositional differences between $n$ = 1.3 and 1.4, an Al content between these values should be the optimal ratio for maximum conversion to MoAlB. Again, excess Al also would compensate its loss due to the high-temperature volatilization. Notably, this proportion with $n$ = 1.2-1.4 is consistent with other studies synthesizing predominantly single-phase MoAlB [28, 29, 44]. In the present work, an Al content of $n$ = 1.3 was used in all subsequent experiments.

Figure 3 shows the product morphology for Mo/1.3Al/B starting powder mixture by SHS. An estimation of the particle size of MoAlB by measuring about 50 particles reveals that products are strip-like or irregular, with the particle size of the length from 3 to 5 μm and width from 0.6 to 1 μm, similar to that by Liu et al [44]. In addition, the elemental mapping analysis of MoAlB by EDS (Figure 3d-f) shows the uniformly distributed elements Al and Mo in the reunited block, where their total atomic ratio much approximates to 1:1 (1:1.06). Notably, a small amount of Mo-Al intermetallic compound is attached to the surface of MoAlB, perhaps resulting in the atomic ratio of Mo to Al not being exactly 1:1.

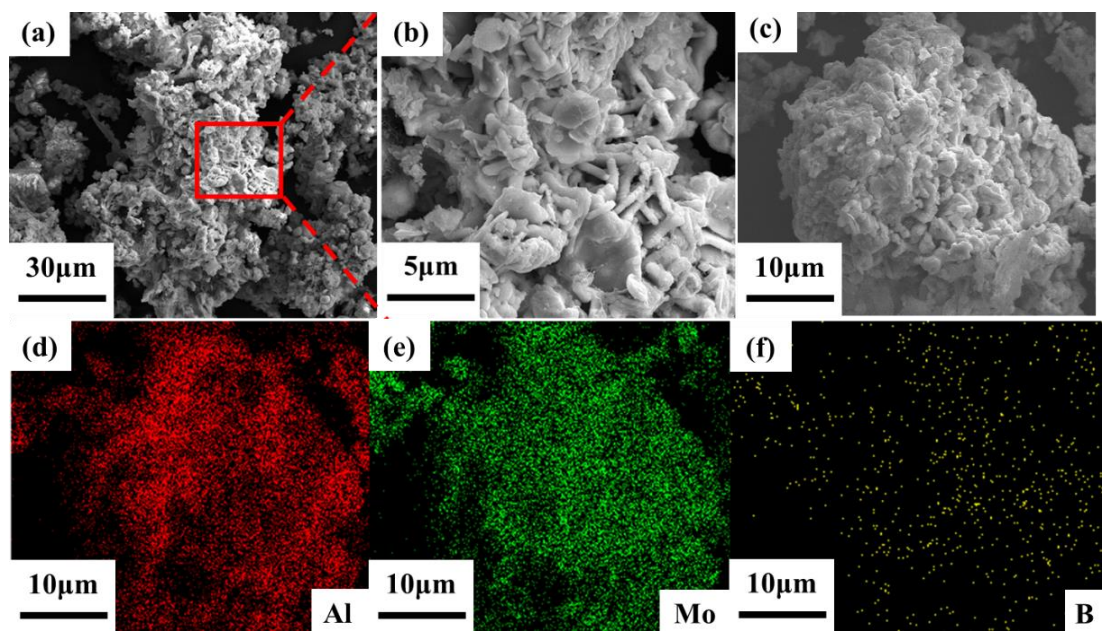

Figure 3 (a) (b) (c) SEM images of samples synthesized from 1Mo/1.3Al/1B by SHS, (b) is the higher magnification micrograph of the particle of red area in (a). Elemental mappings of (d) Molybdenum, (e) Aluminum, and (f) Boron in (c).

*3.1.2 Effect of B molar ratio on formation of MoAlB*

To determine whether the B ratio can affect the content of MoAlB by SHS, the samples of Mo/1.3Al/$n$B ($n$ = 0.95-1.05) raw materials were ignited, with the corresponding XRD patterns in Figure 4 and quantitative phase compositions in Table 1. Unlike the case of Al above, increasing or decreasing the B content cannot improve the purity of the target MoAlB. Moreover, increasing the content of B not only increased the content of $Al_8Mo_3$ and MoB, but also produced a new impurity $MoB_2$ (Figure 4c). Previous studies showed that the decomposition temperature of $MoB_2$ was around 1775 K [45], and SHS occurred at least above 1800 K, indicating that the increase of B may reduce the reaction temperature.

Due to the formation of Mo-Al intermetallic compounds in all situations, and considering that the addition of TiC can improve the purity of $Ti_3AlC_2$ in SHS mode [46], the addition of MoB to the starting materials could also be an effective way to improve the purity of MoAlB. However, after adding only MoB with an atomic ratio of 0.05, ignition cannot be achieved. We speculate that $T_{ad}$ of MoAlB is only slightly higher than 1800 K (this point is verified in the following thermodynamic analysis), which indicates that the heat release during SHS is low except for maintaining basic reaction. However, the reaction between binary borides and Al to produce MoAlB has a higher Gibbs free energy than simple-substance one [31], which means that more external energy input is required.

Table 1 Summary of composition and processing conditions of MoAlB samples using Mo/Al/B powder mixture as starting materials.

| Particle size | Molar ratio | MoAlB | MoB | $Al_8Mo_3$ | $MoB_2$ |
|---|---|---|---|---|---|
| Mo, 10μm; B, 5μm (Coarse raw powder) | 1:1.1:1 | 82% | 9% | 9% | - |
|  | 1:1.2:1 | 87% | 9% | 4% | - |
|  | 1:1.3:1 | 90% | 4% | 6% | - |
|  | 1:1.4:1 | 88% | 2% | 10% | - |
|  | 1:1.3:0.95 | 70% | 7% | 23% | - |
|  | 1:1.3:1.05 | 70% | 5% | 17% | 8% |
| Mo, 1μm; B, 1μm (Fine raw powder) | 1:1.3:1 | 76% | 9% | 16% | - |
| Mo, 500nm; B, 200nm (Ultra-Fine raw powder) | 1:1.3:1 | 74% | 11% | 14% | - |

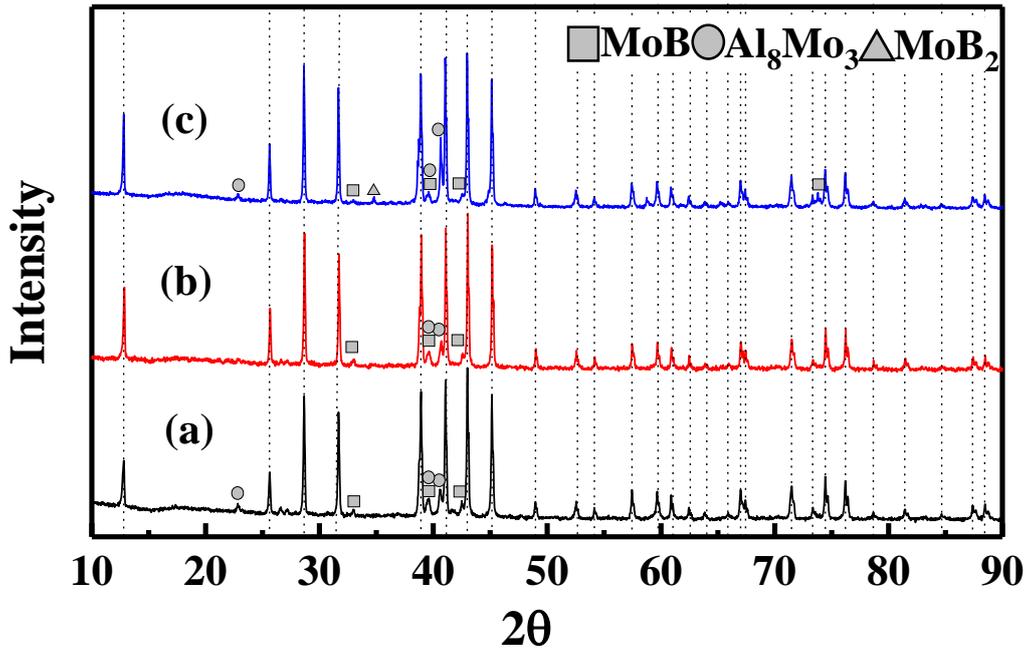

Figure 4 XRD patterns of samples synthesized from Mo/1.3Al/$n$B by SHS for (a) $n$ = 0.95, (b) $n$ = 1, (c) $n$ = 1.05, where the dotted line represents MoAlB

*3.1.3 Effect of raw material particle size on formation of MoAlB*

According to the research of Jessica Merz et al. on the SHS of $Mn_2AlB_2$ [40], the decrease of particle size of raw materials can reduce the ignition temperature of SHS, accelerate the reaction speed, and improve the purity of the product to a certain extent. Therefore, this strategy is also used in the present work on MoAlB. In practice, the Mo/1.3Al/B starting materials with coarse, fine, ultra-fine raw powders were reacted, with their detailed particle size in Table 1 and XRD in Figure 5. Because Al will be completely melted before the reaction, only the change of particle size of Mo and B raw powder is studied here. Still, the main phase MoAlB is detected, with a small amount of impurity phases MoB and $Al_8Mo_3$ (Figure 5). Unlike $Mn_2AlB_2$ [40], the purity of MoAlB does not increase with decreasing the particle size of the raw material.

Figure 6 is morphology of samples from 1Mo/1.3Al/B by SHS for different particle size of raw powder. Of some interest, the plate-like particle is only observed in the one from the ultra-fine raw powders (Figure 6c). This indicates that although changing the particle size of the raw material had a negative impact on the purity of MoAlB, but has an effect on its morphology.

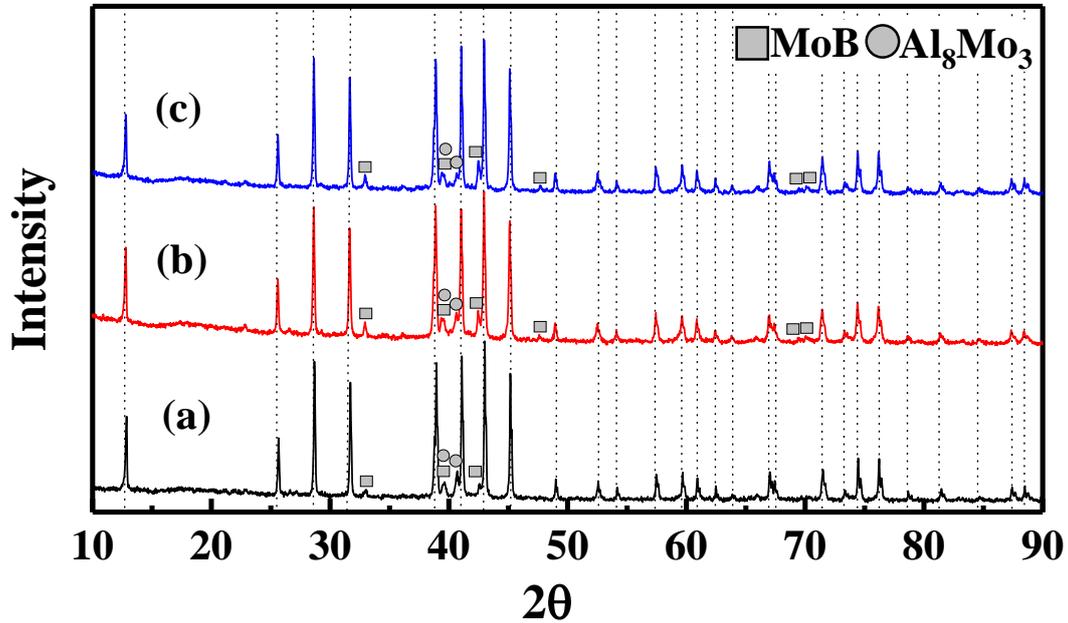

Figure 5 XRD patterns of samples synthesized from Mo/1.3Al/B by SHS with (a) Coarse raw powder, (b) Fine raw powder, and (c) Ultra-Fine raw powder

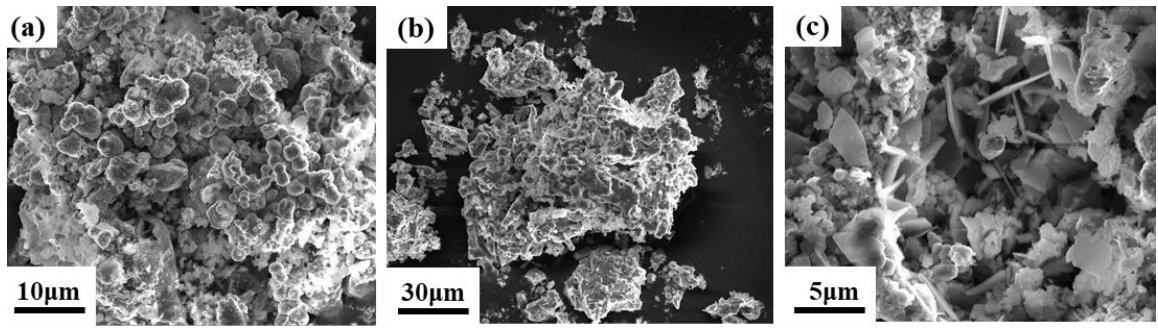

Figure 6 Morphology of samples synthesized from 1Mo/1.3Al/B by SHS for (a) Coarse raw powder, (b)Fine raw powder, and (c) Ultra-Fine raw powder.

Based on study of the thermal explosion reaction mechanism for MoAlB by Liang et al. [31], MoAlB is mainly produced in the following ways:

$$Mo + Al + B = MoAlB \tag{5}$$

$$\text{MoB} + \text{Al} = \text{MoAlB} \tag{6}$$

$$\text{AlB}_2 + 2\,\text{Mo} = \text{MoAlB} + \text{MoB} \tag{7}$$

$$\text{Al}_8\text{Mo}_3 + 3\,\text{B} = 3\,\text{MoAlB} + 5\,\text{Al} \tag{8}$$

$$3\,\text{Mo}_2\text{B} + 11\,\text{Al} = 3\,\text{MoAlB} + \text{Al}_8\text{Mo}_3 \tag{9}$$

Based on the existing experimental results, with the increase of Al content in the raw material, the ratio of MoAlB and $Al_8Mo_3$ increases and the ratio of MoB decreases, indicating that reactions (6) and (9) should be the two main reaction pathways in the SHS process, but it should be noted that reaction (6) can only partially improve the purity of the product. The product does not contain $Mo_2B$ phase, which could be attributed to the fact its decomposition temperature is lower than $T_{ad}$. As a result, $Mo_2B$ has never been found in previous SHS research of Mo-B system [47-50]. However, changing the ratio of B did not improve the purity of the product, implying that reaction (8) does not play a major role in the SHS process. According to reaction (7), although increasing Mo can increase the content of MoAlB, it will also increase the amount of MoB, which does not improve the purity of the product.

*3.2 Thermodynamic analysis of SHS reaction*

The thermodynamic analysis of SHS reaction previously described is usually expressed in the form of an 'enthalpy vs temperature' curve [51]. While data exist for the specific heat capacities of MoAlB [19] but the data of enthalpy of formation are limited. Sankalp Kota et al. calculated thermochemical data of MoAlB phases based on Gibbs free-energy of MoB and Al, and have reported -132±3.2 kJ/mol for the enthalpy of formation of MoAlB at 298 K [19]. In Table 2, all the thermodynamic data of products and the reactants (Mo, Al, B) were then used for the follow-up analysis [19, 52].

Table 2 Enthalpy of formation ($\Delta H_f$, kJ/mol), fusion ($\Delta H_m$, kJ/mol), and specific heat capacities ($C_p$, J K$^{-1}$ mol$^{-1}$) with temperature of reactant and product, $T$ (K).

| Compound | $C_p = a+bT+cT^{-2}+dT^2$ (J·K$^{-1}$·mol$^{-1}$) | | | | Temperature range (K) | Status | $\Delta H_f$ (kJ/mol) | $\Delta H_m$ (kJ/mol) |
|---|---|---|---|---|---|---|---|---|
| | $a$ | $b$ ($\times 10^{-3}$) | $c$ ($\times 10^5$) | $d$ ($\times 10^{-6}$) | | | | |
| Mo | 25.568 | 2.845 | -2.184 | - | 298.15-700 | s | - | - |
| | 33.911 | -11.912 | -9.205 | 6.958 | 700-1500 | s | - | - |
| | 16.669 | 9.694 | - | - | 1500-2000 | s | - | - |
| | 206.347 | -126.620 | -1053.799 | 27.338 | 2000-2892 | s | - | - |
| Al | 31.376 | -16.393 | -3.607 | 20.753 | 298.15-933 | s | - | 10.71 |
| | 31.748 | - | - | - | 933-2767 | l | - | |
| B | 27.815 | -0.699 | -32.171 | 5.209 | 298.15-800 | s | - | - |
| | 21.372 | 4.72 | -12.108 | - | 800-1500 | s | - | - |
| | 32.095 | 0.071 | -96.751 | - | 1500-2450 | s | - | - |
| MoAlB | 72.363 | 8.73 | -13.45 | -0.98 | 298.15-2000 | s | -132 ± 3.2 | - |

*3.2.1 Enthalpy-temperature curves for the Mo–Al–B system*

Figure 7 is the enthalpy-temperature curves, which plot the enthalpy associated with heating the reactants and products. In left side of Figure 7, enthalpy of the reaction ($\Delta H_{AB}^{298}$) represents the left side of reaction (2) (Sections A-B). By creating a horizontal line to the enthalpy of the products, $T_{ad}$ value (1979 K) can be found from the vertical line to horizontal axis, which represents the right hand side of the reaction (2) ($y = 0$) (Sections C-D). According to the research of Kota et al. [19], the decomposition temperature of MoAlB is around 1708 K, and $T_{ad}$ is greater than 1708 K, which means that MoAlB should formed in the cooling of the products after SHS. With the gradual increase of Al from 0 to 0.4 ($y$), the values of $T_{ad}$ are 1979 K, 1909 K, 1848 K, 1788 K, and 1728 K, respectively. Clearly, excess Al as a diluent result in a decrease of $T_{ad}$, but offers the potential to control the exothermicity of the SHS.

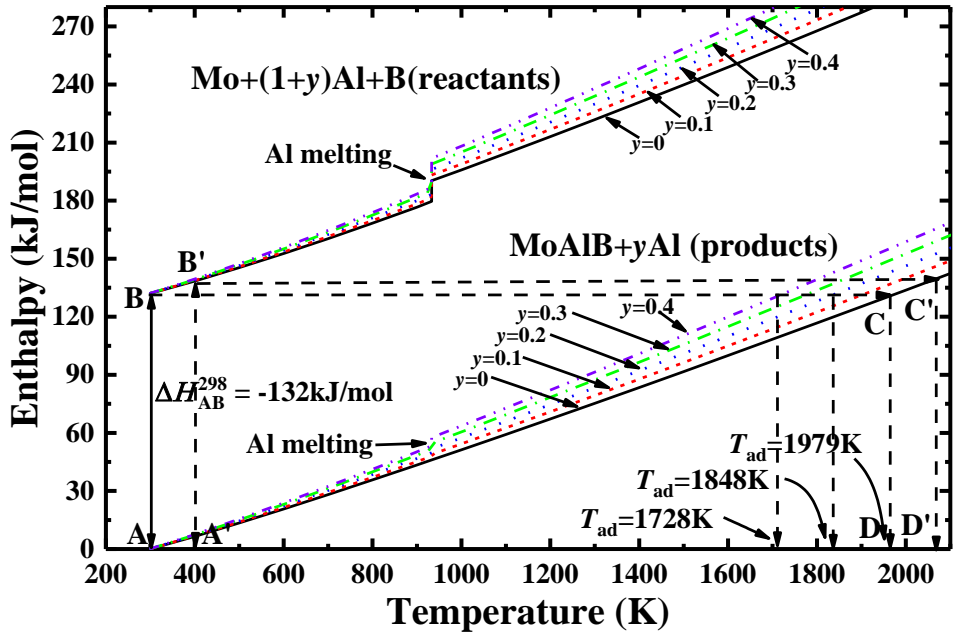

Figure 7 Enthalpy-temperature SHS analysis of MoAlB + (1+ y) Al + B = MoAlB + yAl. ABCD and A'B'C'D' represent the calculation process of the adiabatic temperature from 1Mo/1Al/1B precursors at room temperature and preheated to 400K, respectively.

However, it can be found that $T_{ad}$ decreases below 1800 K when increasing y at and above 0.3, different from the experimental ones that they can still be ignited until y = 0.4, which indicates that the present calculated $T_{ad}$ is slightly lower than measured one. This is also different from the previous rule that the calculated temperature of adiabatic temperature in MAX phase is often higher than the actual measured temperature duo to heat losses [53, 54]. The enthalpy in work is based on the MoAlB compound containing $Al_2O_3$ and $Al_3Mo$ impurities [19], which may lead to the lower value than its real enthalpy of pure MoAlB and the calculated adiabatic temperature decreased. To obtain accurate prediction, the future work is to use more accurate thermodynamic properties of pure MoAlB.

If the reactants are preheated prior to SHS, this additional energy is present to increase $T_{ad}$. Taking the pre-heating of reactants to 400 K as an example, considering the heat absorbed by the reactants in Eq. (4), the calculation process corresponds to Section A'B'C'D' in Figure 7. Here, the corresponding adiabatic temperature is increased to 2054 K. Figure 8 summaries the change in $T_{ad}$ with pre-heating temperature for $Mo + Al + B = MoAlB$.

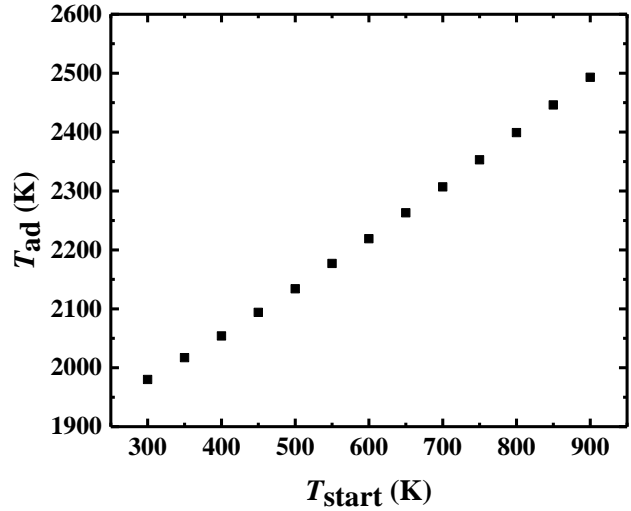

Figure 8 $T_{ad}$ as a function of the starting temperatures ($T_{start}$) for the SHS reaction of $Mo + Al + B = MoAlB$.

*3.2.2 MoAlB satisfying condition for SHS*

The manufacture of MoAlB by SHS has been examined by experimental results and the reaction is self-sustaining, therefore it is possible to assess whether two relationships, (i) $T_{ad} \geqslant 1800\,K$ [55] and (ii) $-\Delta H_r^{298K}/C_p^{298K} \geqslant 2000\,K$ [56], hold true for this particular MAB phase system.

From the thermodynamic analysis, the $T_{ad}$ is 1979 K (when $y = 0$) and therefore satisfies condition (i). For condition (ii), since $-\Delta H_r^{298K} = 132$ kJ/mol and $C_p^{298K} = 58.46$ J/(mol·K), the ratio $-\Delta H_r^{298K}/C_p^{298K} = 2258$ K and also exceeds 2000 K. Figure 9 shows the relatively linear relationship between $-\Delta H_r^{298K}/C_p^{298K}$ and $T_{ad}$ for a range of selected compounds in the SHS literature [41]. It is interesting to note that the MAB phase reactions considered in this work also follow this relationship. These two relatively simple criteria for predicting SHS therefore work for this particular MAB phase system and may also be applicable to the many other MAB phases family in the next work.

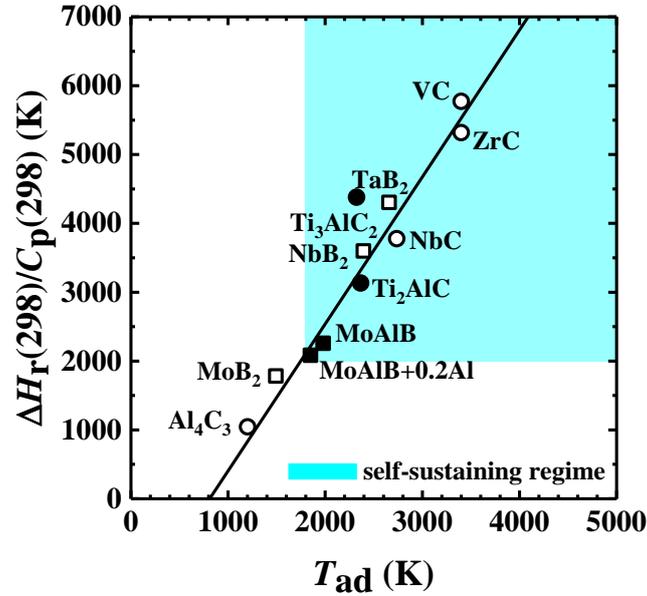

Figure 9 Relationship between $\Delta H_r(298)/C_p(298)$ (K) and $T_{ad}$ for selected compounds. Most of Compounds are reported in [41] . Compounds (■) (□) are the MAB phase and binary borides, Compounds (●) (○) are the MAX phase and binary carbide. The self-sustaining regime on the right side indicates that the material can be synthesized using the SHS method with $T_{ad} \geq 1800$ K and $-\Delta H_r^{298K}/C_p^{298K} \geq 2000$ K.

4. Conclusions

Ternary-layered transition-metal boride MoAlB is successfully prepared by self-propagating high temperature combustion synthesis in a ratio of 1Mo/1.3Al/1B by SHS within a few seconds. The increase of Al content below 1.3 can improve the purity of MoAlB, while the change of the B content and the particle size of the raw material cannot improve the purity of the production, but the decrease of the particle size of the raw material may promote the transformation of the particle morphology from granular-like to flake-like. A thermodynamic analysis of MoAlB shows that the adiabatic temperature decreases with increasing Al content and decreasing pre-heating temperature. The successful validation of a series of SHS reaction conditions in MoAlB ceramics offers the criteria of predicting large-scale preparation of other high-cost MAB phases.

**Declaration of competing interest**

The authors have no competing interests to declare that are relevant to the content of this article.

**Acknowledgment**

This work was supported by the National Natural Science Foundation of China (Grant No.51972080), National Key Research and Development Program of China (Grant No.7 2018YFA0702802), and Shenzhen Science and Technology Program.

**References**

[1] S. You, Z. Yu, Z. Yang, Research progress of boride ceramics and their composite materials, Physical Testing and Chemical Analysis Part B: Physical Testing and Chemical Analysis Part one, 3 (2007) 27-31.

[2] Y. Bai, N. Srikanth, C.K. Chua, K. Zhou, Density functional theory study of $M_{n+1}AX_n$ phases: a review, Critical Reviews in Solid State & Material Sciences, In press (2017) 1-51.

[3] X. Qi, W. Yin, S. Jin, A. Zhou, X. He, G. Song, Y. Zheng, Y. Bai, DFT predictions and experimental confirmation of mechanical behaviour and thermal properties of the

Ga-bilayer $Mo_2Ga_2C$, Journal of Advanced Ceramics, 11 (2021) 273-282.

[4] A. Zhou, Y. Liu, S. Li, X. Wang, G. Ying, Q. Xia, P. Zhang, From structural ceramics to 2D materials with multi-applications: A review on the development from MAX phases to MXenes, Journal of Advanced Ceramics, 10 (2021) 1194-1242.

[5] M. Ade, H. Hillebrecht, Ternary borides $Cr_2AlB_2$, $Cr_3AlB_4$, and $Cr_4AlB_6$: the first members of the series $(CrB_2)_nCrAl$ with n=1, 2, 3 and a unifying concept for ternary borides as MAB-phases, Inorg. Chem., 54 (2015) 6122-6135.

[6] Y. Bai, X. Qi, A. Duff, N. Li, F. Kong, X. He, R. Wang, W.E. Lee, Density functional theory insights into ternary layered boride MoAlB, Acta Materialia, 132 (2017) 69-81.

[7] Y.L. Bai, H. Yin, G.P. Song, X.D. He, X.X. Qi, J. Gao, B.B. Hao, J.Z. Zhang, High-fracture-toughness ternary layered ceramics: from the MAX to MAB phases, Journal of Materials Engineering, 49 (2021) 1-23.

[8] S. Kota, M. Sokol, M.W. Barsoum, A progress report on the MAB phases: atomically laminated, ternary transition metal borides, International Materials Reviews, 65 (2019) 226-255.

[9] W. Zhang, S. Li, S. Wu, B. Yao, S. Fan, G. Bei, W. yu, Y. Zhou, Y. Wu, S.-A. Ding, Synthesis and properties of MoAlB composites reinforced with SiC particles, Journal of Advanced Ceramics, 11 (2022) 1-9.

[10] W. Jeitschko, The crystal structure of $Fe_2AlB_2$, Acta Crystallographica, 25 (1969) 163-165.

[11] H. Nowotny, P. Rogl, Ternary Metal Borides, Boron and Refractory Borides, Springer, Berlin, Heidelberg, 1977, pp. 413-438.

[12] X. Tan, P. Chai, C.M. Thompson, M. Shatruk, Magnetocaloric effect in $AlFe_2B_2$: toward magnetic refrigerants from earth-abundant elements, Journal of the American Ceramic Society, 135 (2013) 9553-9557.

[13] Y. Bai, X. Qi, X. He, D. Sun, F. Kong, Y. Zheng, R. Wang, A.I. Duff, Phase stability and weak metallic bonding within ternary-layered borides CrAlB, $Cr_2AlB_2$,$Cr_3AlB_4$,and $Cr_4AlB_6$, Journal of the American Ceramic Society, 102 (2018) 3715-3727.

[14] S. Kota, E. Zapata-Solvas, A. Ly, J. Lu, O. Elkassabany, A. Huon, W.E. Lee, L. Hultman, S.J. May, M.W. Barsoum, Synthesis and Characterization of an Alumina Forming Nanolaminated Boride: MoAlB, Scientific reports, 6 (2016) 1-11.

[15] N. Li, Y. Bai, S. Wang, Y. Zheng, F. Kong, X. Qi, R. Wang, X. He, A.I. Duff, Rapid synthesis, electrical, and mechanical properties of polycrystalline $Fe_2AlB_2$ bulk from elemental powders, Journal of the American Ceramic Society, 100 (2017) 4407-4411.

[16] Y. Bai, D. Sun, N. Li, F. Kong, X. Qi, X. He, R. Wang, Y. Zheng, High-temperature mechanical properties and thermal shock behavior of ternary-layered MAB phases $Fe_2AlB_2$, International Journal of Refractory Metals and Hard Materials, 80 (2019) 151-160.

[17] L. Xu, O. Shi, C. Liu, D. Zhu, S. Grasso, C. Hu, Synthesis, microstructure and properties of MoAlB ceramics, Ceramics International, 44 (2018) 13396-13401.

[18] Y. Bai, X. Qi, X. He, G. Song, Y. Zheng, B. Hao, H. Yin, J. Gao, A. Ian Duff, Experimental and DFT insights into elastic, magnetic, electrical, and thermodynamic properties of MAB-phase $Fe_2AlB_2$, Journal of the American Ceramic Society, 103 (2020) 5837-5851.

[19] S. Kota, M. Agne, E. Zapata Solvas, O. Dezellus, D. Lopez, B. Gardiola, M. Radovic, M.W. Barsoum, Elastic properties, thermal stability, and thermodynamic parameters of MoAlB, Physical Review B, 95 (2017) 114108.

[20] S. Kota, E. Zapata-Solvas, Y. Chen, M. Radovic, W.E. Lee, M.W. Barsoum, Isothermal and Cyclic Oxidation of MoAlB in Air from 1100°C to 1400°C, Journal of

the Electrochemical Society, 164 (2017) C930-C938.

[21] X. Lu, S. Li, W. Zhang, B. Yao, W. Yu, Y. Zhou, Crack healing behavior of a MAB phase: MoAlB, Journal of the European Ceramic Society, 39 (2019) 4023-4028.

[22] G. Bei, S. van der Zwaag, S. Kota, M.W. Barsoum, W.G. Sloof, Ultra-high temperature ablation behavior of MoAlB ceramics under an oxyacetylene flame, Journal of the European Ceramic Society, 39 (2019) 2010-2017.

[23] Z. Guo, J. Zhou, Z. Sun, New two-dimensional transition metal borides for Li ion batteries and electrocatalysis, Journal of Materials Chemistry A, 5 (2017) 23530-23535.

[24] M. Jakubczak, A. Szuplewska, A. Rozmysłowska-Wojciechowska, A. Rosenkranz, A.M. Jastrzębska, Novel 2D MBenes-Synthesis, Structure, and Biotechnological Potential, Advanced Functional Materials, 31 (2021) 2103048.

[25] L.T. Alameda, P. Moradifar, Z.P. Metzger, N. Alem, R.E. Schaak, Topochemical deintercalation of Al from MoAlB: Stepwise etching pathway, layered Intergrowth structures, and two-dimensional MBene, Journal of the American Chemical Society, 140 (2018) 8833-8840.

[26] P. Helmer, J. Halim, J. Zhou, R. Mohan, B. Wickman, J. Björk, J. Rosén, Investigation of 2D Boridene from First Principles and Experiments, Advanced Functional Materials, (2022) 2109060.

[27] J. Zhou, J. Palisaitis, J. Halim, M. Dahlqvist, Q. Tao, I. Persson, L. Hultman, P.O.Å. Persson, J. Rosen, Boridene: Two-dimensional $Mo_{4/3}B_{2-x}$ with ordered metal vacancies obtained by chemical exfoliation, Science, 373 (2021) 801.

[28] E. Wang, Y. Guo, C. Guo, T. Yang, X. Hou, Z. He, H. Wang, Effect of temperature on the initial reaction behavior of MAB phases (MoAlB powders) at 700–1000°C in air, Ceramics International, 47 (2021) 20700-20705.

[29] S. Wang, Y. Xu, Z. Yu, H. Tan, S. Du, Y. Zhang, J. Yang, W. Liu, Synthesis,

microstructure and mechanical properties of a MoAlB ceramic prepared by spark plasma sintering from elemental powders, Ceramics International, 45 (2019) 23515-23521.

[30] X. Su, J. Dong, L. Chu, H. Sun, S. Grasso, C. Hu, Synthesis, microstructure and properties of MoAlB ceramics prepared by in situ reactive spark plasma sintering, Ceramics International, 46 (2020).

[31] B. Liang, Z. Dai, W. Zhang, Q. Li, D. Niu, M. Jiao, L. Yang, X. Guan, Rapid synthesis of MoAlB ceramic via thermal explosion, Journal of Materials Research and Technology, 14 (2021) 2954-2961.

[32] A. Merzhanov, I. Borovinskaya, Historical retrospective of SHS: An autoreview, International Journal of Self-Propagating High-Temperature Synthesis, 17 (2008) 242-265.

[33] Y. Bai, X. He, Y. Li, C. Zhu, S. Zhang, Rapid synthesis of bulk $Ti_2AlC$ by self-propagating high temperature combustion synthesis with a pseudo-hot isostatic pressing process, Journal of Materials Research, 24 (2009) 2528-2535.

[34] Y. Bai, X. He, R. Wang, Y. Sun, C. Zhu, S. Wang, G. Chen, High temperature physical and mechanical properties of large-scale $Ti_2AlC$ bulk synthesized by self-propagating high temperature combustion synthesis with pseudo hot isostatic pressing, Journal of the European Ceramic Society, 33 (2013) 2435-2445.

[35] J.M. Guo, K.X. Chen, Z.B. Ge, H.P. Zhou, X.S. Ning, Effects of different Ti/C and Al contents on combustion synthesized $Ti_3AlC_2$ powders in the Ti-Al-C system, Rare Metal Materials and Engineering, 32 (2003) 561-565.

[36] Z. Ge, K. Chen, J. Guo, H. Zhou, J.M.F. Ferreira, Combustion synthesis of ternary carbide $Ti_3AlC_2$ in Ti-Al-C system, Journal of the European Ceramic Society, 23 (2003) 567-574.

[37] C. Yeh, Y. Shen, Effects of SiC addition on formation of $Ti_3SiC_2$ by self-propagating high-temperature synthesis, Journal of Alloys and Compounds, 461 (2008) 654-660.

[38] Y. Li, P. Bai, The microstructural evolution of $Ti_2SnC$ from Sn-Ti-C system by Self-propagating high-temperature synthesis (SHS), International Journal of Refractory Metals and Hard Materials, 29 (2011) 751-754.

[39] C.W. Yeh, C.W. Kuo, An investigation on formation of $Nb_2AlC$ by combustion synthesis of $Nb_2O_5$-Al-$Al_4C_3$ powder compacts, Journal of Alloys and Compounds, 496 (2010) 566-571.

[40] J. Merz, P. Richardson, D. Cuskelly, Formation of $Mn_2AlB_2$ by Induction-Assisted Self-Propagating High-Temperature Synthesis, Open Ceramics, 8 (2021) 100190.

[41] T. Thomas, C.R. Bowen, Thermodynamic predictions for the manufacture of $Ti_2AlC$ MAX-phase ceramic by combustion synthesis, Journal of Alloys and Compounds, 602 (2014) 72-77.

[42] S. Chen, L. Wang, G. He, J. Li, C.-A. Wang, Microstructure and properties of porous Si3N4 ceramics by gelcasting-self-propagating high-temperature synthesis (SHS), Journal of Advanced Ceramics, 11 (2022) 172-183.

[43] K. Morsi, The diversity of combustion synthesis processing: a review, Journal of Materials Science, 47 (2012) 68-92.

[44] C. Liu, Z. Hou, Q. Jia, X. Liu, S. Zhang, Low Temperature Synthesis of Phase Pure MoAlB Powder in Molten NaCl, Materials, 13 (2020) 758.

[45] T.B. Massalski, H. Okamoto, P. Subramanian, L. Kacprzak, W.W. Scott, Binary alloy phase diagrams, OH: American society for metals, Metals Park, 1986.

[46] H.Y. Sun, X. Kong, Z.Z. Yi, Q.B. Wang, G.Y. Liu, The difference of synthesis mechanism between $Ti_3SiC_2$ and $Ti_3AlC_2$ prepared from Ti/M/C (M = Al or Si)

elemental powders by SHS technique, Ceramics International, 40 (2014) 12977-12981.

[47] C.L. Yeh, W.S. Hsu, Preparation of MoB and MoB-MoSi$_2$ composites by combustion synthesis in SHS mode, Journal of Alloys and Compounds, 440 (2007) 193-198.

[48] A.Y. Potanin, S. Vorotilo, Y.S. Pogozhev, S.I. Rupasov, T.A. Lobova, E.A. Levashov, Influence of mechanical activation of reactive mixtures on the microstructure and properties of SHS-ceramics MoSi$_2$-HfB$_2$-MoB, Ceramics International, 45 (2019) 20354-20361.

[49] S. Vorotilo, A.Y. Potanin, Y.S. Pogozhev, E.A. Levashov, N.A. Kochetov, D.Y. Kovalev, Self-propagating high-temperature synthesis of advanced ceramics MoSi$_2$–HfB$_2$–MoB, Ceramics International, 45 (2019) 96-107.

[50] E.G. Lavut, N.V. Chelovskaya, O.E. Kashireninov, Direct determination of the enthalpy of formation of MoB in synthesis from simple substances in an SHS system, Journal of Engineering Physics and Thermophysics, 65 (1993) 971-973.

[51] J. Rodrigues, V. Pandolfelli, W. Botta, R. Tomasi, B. Derby, R. Stevens, R. Brook, Thermodynamic Predictions for the Formation of Ceramic-Metal Composite by Self-Propagating High-Temperature Synthesis, Journal of Materials Science Letters, 10 (1991) 819-823.

[52] M.E. Brown, P.K. Gallagher, Handbook of thermal analysis and calorimetry: applications to inorganic and miscellaneous materials, Elsevier, 2003.

[53] S. Hashimoto, N. Nishina, K. Hirao, Y. Zhou, H. Hyuga, S. Honda, Y. Iwamoto, Formation mechanism of Ti$_2$AlC under the self-propagating high-temperature synthesis (SHS) mode, Materials Research Bulletin, 47 (2012) 1164-1168.

[54] A. Stolin, D. Vrel, S. Galyshev, A. Hendaoui, P. Bazhin, A. Sytschev, Hot forging of MAX compounds SHS-produced in the Ti-Al-C system, International journal of self-

propagating high-temperature synthesis, 18 (2009) 194-199.

[55] A.G. Merzhanov, The chemistry of self-propagating high-temperature synthesis, Journal of Materials Chemistry, 14 (2004) 1779-1786.

[56] L.L. Wang, Z. Munir, Y.M. Maximov, Thermite reactions: their utilization in the synthesis and processing of materials, Journal of Materials Science, 28 (1993) 3693-3708.